\documentclass{IEEEcsmag}
\usepackage[colorlinks,urlcolor=blue,linkcolor=blue,citecolor=blue]{hyperref}
\expandafter\def\expandafter\UrlBreaks\expandafter{\UrlBreaks\do\/\do\*\do\-\do\~\do\'\do\"\do\-}
\usepackage{upmath,color}

\usepackage[utf8]{inputenc} %
\usepackage[T1]{fontenc}

\usepackage[super,comma,compress]{natbib}

\usepackage{hyperref}

\usepackage{multirow}
\usepackage{booktabs}
\usepackage{tabularx}
\usepackage{arydshln}
\usepackage{todonotes}
\usepackage{makecell}
\usepackage{graphicx}

\usepackage{xspace}

\newcommand{\eg}{\textit{e.g.,}~}
\newcommand{\ie}{\textit{i.e.,}~}
\newcommand{\cf}{\textit{cf.,}\xspace}

\newcommand{\one}{({\em i})\xspace}
\newcommand{\two}{({\em ii})\xspace}
\newcommand{\three}{({\em iii})\xspace}

\makeatletter
\renewcommand{\paragraph}[1]{\vspace*{0.03in}\noindent{\bf #1.}\hspace{0.25ex \@plus1ex \@minus.2ex}}
\newcommand{\paragraphNoDot}[1]{\vspace*{0.03in}\noindent{\bf #1}\hspace{0.25ex \@plus1ex \@minus.2ex}}
\makeatother

\newcommand{\caaTag}[1]{\texttt{#1}~}
\newcommand{\tagI}{\caaTag{issue}}
\newcommand{\tagIW}{\caaTag{issuewild}}

\begin{document}

\title{A Call to Reconsider \\ Certification Authority Authorization (CAA)}

\author{Pouyan Fotouhi Tehrani}
\affil{TU Dresden, Germany}

\author{Raphael Hiesgen}
\affil{HAW Hamburg, Germany}

\author{Thomas C. Schmidt}
\affil{HAW Hamburg, Germany}

\author{Matthias W\"ahlisch}
\affil{TU Dresden, Germany}

\begin{abstract}
Certification Authority Authentication (CAA) is a safeguard against illegitimate certificate issuance. We show how shortcomings in CAA concepts and operational aspects undermine its effectiveness in preventing certificate misissuance. Our discussion reveals pitfalls and highlights best practices when designing security protocols based on DNS.

\vspace*{5pt}
\textbf{Keywords: Web PKI, DNS, CAA, Protocol Security}
\end{abstract}

\maketitle

\section{Introduction}
\label{sec:intro}
 Web security~\cite{tosw-smwc-24} relies on X.509 certificates.
  Fundamental to the underlying security model is the correct issuance of certificates.
This trust is challenged by unrestricted certification.
Certification Authorities (CA) are generally allowed to certify any arbitrary resource, \ie to bind a public key to a domain name by creating a certificate.
If certification occurs without the consent of a name owner, the misissued certificate can be used to impersonate resources of that name owner.
Causes for misissuance are manyfold.
Rogue or compromised CAs~\cite{agsbc-mahsa-17}, spoofed DNS records or compromised name servers~\cite{sww-difsd-19}, and malicious traffic rerouting~\cite{g-bhwbh-15,bserm-hyajp-2018} have been part of attack vectors in the past.

There are two principal directions to harden the security of the certificate issuance process.
\one \emph{Prevention} by enabling a CA to verify whether the issuing request is valid.
\two
\emph{Mitigation} by enabling the name owner or any third party on behalf to identify and revoke an incorrectly issued certificate.
Both directions have been covered in standardization and deployment.
Certificate transparency (CT)~\cite{RFC-9162} makes certificates public and DNS-Based Authentication of Named Entities for TLS (DANE TLSA)~\cite{RFC-6698} binds certificates to services available below a specific domain name.
Prevention is specified in CA authorization (CAA)~\cite{RFC-8659}, which allows a name owner to note in the DNS which CA is allowed to issue a certificate.

In this paper, we argue that both prevention and 
mitigation are important, but that prevention should be the first-class citizen because it addresses a root cause~(\autoref{sec:background}).
Unfortunately, CAA, which is not only an IETF standard but also the solution the CA/Browser Forum agreed on to be mandatory, is flawed.
We revisit design decisions and deployment behaviors (\autoref{sec:op-pitfalls}) to refine on-going standards and inform our community to guide fundamentally different approaches~(\autoref{sec:disc}), hoping that these insights prevent common pitfalls in the future. 
Our data-driven approach (\autoref{sec:method}) is based on more than $4.6M$ unique certificates from CT logs, which we test whether they conform to CAA policies.
Our major findings read:
\begin{enumerate}
    \item Implicit semantics overshadow expressiveness.
    \item Underspecified syntax allows misinterpretation.
    \item Loose policy scoping raises security risks.
    \item Misaligned procedures defeat reliability and trust.
\end{enumerate}

\section{Background}
\label{sec:background}
CAA~\cite{RFC-8659} is a DNS-based approach that allows a domain name owner to specify which CA is allowed to issue certificates for the namespace under her control.
Dedicated \texttt{CAA} DNS resource records contain this information.
Checking CAA records 
was made mandatory for CAs by the CA/Browser Forum in 2017.

When a CA verifies a certificate issue request for a domain name, the CA needs to perform two steps.
First, the CA needs to find DNS material that is considered for CA authorization.
Second, based on the obtained DNS material, the CA checks whether it is authorized to issue a certificate for the domain name.

A CA starts searching for DNS material at the given domain name (\eg www.example.com). If no CAA records are found, the CA iteratively traverses the parent hierarchy (\eg example.com, then com) until CAA records are found or the root zone (\texttt{.}) is reached.
For aliases, the canonical name (denoted by a \texttt{CNAME} record) is first checked and if no CAA records are found, the parents of the alias
 are visited.

A CA can verify whether it is authorized to issue the requested certificate based on the CAA model, which provides properties (also called \emph{tag}) and values as part of a CAA record.
The IETF standardized \texttt{issue} and \texttt{issuewild} to constrain issuance and \texttt{iodef} to contact the name owner in case of policy violations.
The value of \tagI and \tagIW contains 
an `issuer-domain-name' to identify a CA and a list of optional parameters.
CAs choose their own identifiers and list them in their Certification Practice Statement (CPS).
It is possible to constrain issuance to multiple CAs or completely forbid issuance by having an \tagI or \tagIW without any valid CA identifier. 
The \caaTag{iodef} value must be a URL using \texttt{mailto}, \texttt{http}, or \texttt{https} as scheme.

Two additional tags are defined by the CA/Browser Forum to verify ownership of a name via phone (\texttt{contact\-phone}) or e-mail (\texttt{contactemail}).
The CAA specification also allows non-standard tags to introduce new semantics and functionalities.
If a CA encounters an unsupported tag, it either rejects issuance if its critical flag is set, otherwise ignores the record.
If only non-critical unknown tags or only tags other than \texttt{issue} or \texttt{issuewild} are found,
it is interpreted as no issuance constraint.

CAA does not repel misbehaving CAs as it
cannot prevent records from being ignored.
Third-party Evaluators~\cite{RFC-8659}, however, could audit CAs and degrade CAs in case of misbehavior, possibly leading to a healthier ecosystem in the mid-term.
Nevertheless, we argue that there are further serious concerns because CAA specification~\cite{RFC-8659} and deployment agreements~\cite{bf-brimp-24} suffer from four blind spots:

\begin{enumerate}
    \item {\bf Implicit semantics} mask misconfigurations and reduces expressiveness of CAA policies.
    \item {\bf Boundless policy scoping} allows parent zones to define policies
     if child zones have no explicit policies.
    \item {\bf Ambiguous CA identifiers} can  (unintentionally) lead to overly permissive policies.
    \item {\bf Non-verifiable and temporally unrestricted policies} make CAA-based CA auditing unreliable if not impossible.
\end{enumerate}

\noindent
These four properties practically defeat the primary goal of CAA, \ie constraining issuance, as well as its secondary goal, \ie enabling \emph{Evaluators} to audit CA issuance behavior, as we discuss in the following.

\section{Threat Model}
\label{sec:threat-model}

\begin{figure}
    \centering
    \includegraphics{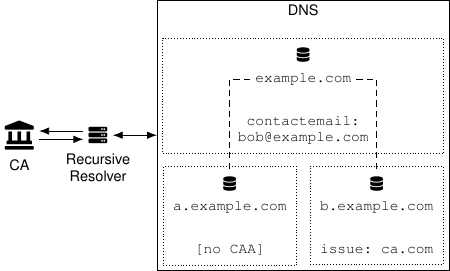}
    \caption{Overview of entities involved in CAA validation in a namespace with two delegations (\texttt{\{a,b\}.example.com}).}
    \label{fig:caa-boundary}
\end{figure}

CAA is based on the unprotected DNS, so it inherits all of its security shortcomings.
For example, if an attacker manages to compromise the name servers of a domain name, the attacker can modify or remove CAA records to undermine security policies.
Furthermore, CAA does not protect against malicious or misbehaving CAs.
In this paper, we consider these aspects as out-of-scope, and assume the following in our threat model: \one target domain name servers and respective DNS records are intact and authoritative, and \two CAs correctly implement and respect CAA.
Actors are name owners, operators of public recursive resolvers, CAs, and an active attacker as adversary.
The attacker cannot manipulate or compromise name owner infrastructure but other name servers and recursive resolvers.
The CA uses a recursive resolver to gather CAA records before issuance (see \autoref{fig:caa-boundary}).
In the following, we discuss in detail why certificate issuance is at risk despite CAA.

\paragraph{Spoofing name owner}
Assume an attacker that aims at manipulating CAA policies
 to authorize or deny a specific CA, or to insert its own contact information for domain validation.
If the domain name has its own CAA policies, an attack is not feasible without compromising the target name servers or manipulating DNS message distribution.
The prior requirement contradicts our assumptions (secure target domain name server, and good CA), the latter could be countered by enabling the (optional) DNSSEC.

If the target domain does not have its own CAA policies, a viable attack vector is to convince or compromise a name server along the path to the root zone, or even take advantage of insecure CAA policies on those name servers.
In such scenario, regardless of security measures taken by a name owner, spoofing is possible.
This follows from CAA specification to search for CAA records beyond a DNS zone boundaries (see \autoref{sec:background}) and does not require CAA policies for a domain name to be defined by its respective owner.
An example is \texttt{a.example.com} in \autoref{fig:caa-boundary}.

A similar case occurs when a name alias points to a canonical name using a \texttt{CNAME} record.
Here, only the owner of the canonical name can define CAA policies for the aliased resources.
The attack vector then transfers to the canonical name, \eg by targeting CDN operators that commonly use canonical names for their customers.

Two possible solutions exist for this issue.
The CAA lookup strategy could be sharpened to remain confined to the actual name owner.
Alternatively, a subdomain could be designated for CAA records that eliminates the recursive CAA lookup algorithm.
We discuss these approaches in depths in \autoref{subsec:problem-boundless}.

\paragraph{Elevation of privilege}
Assume  an attacker that wants to authorize further CAs  for a given domain name beyond those intended by the name owner, \eg to take some advantage of issuance, such as a weak validation procedure.
To reach this goal, an attacker can take advantage of the fact that CA identifiers are not assigned in a coordinated or centralized manner (\eg through IANA), and that the CAA specification does not require uniqueness or exclusive use of identifiers.
Respectively, an attacker could convince a name owner to add or use some identifier which also authorizes another, unintended CA (see \autoref{tab:ca_string_shares}).

This attack depends on the deployment and can be mitigated by \one coordination among CAs when choosing their identifiers, or by \two provision of a public index that makes CA identifiers transparent to name owners.
We discuss this issue in details in \autoref{subsec:design-amb-id}.

\paragraph{Repudiation}
CAs document the issuance procedure together with relevant CAA records and are even mandated to do so in cases where issuance was prevented by a CAA record.
However, in case of a misissuance, these records cannot be used to prove or refute wrongdoing by the CA as DNS records are neither signed nor carry a validity period. It is impossible to assert their authenticity or validity.

If an attacker, for example, manages to compromise DNS resolvers used by a CA, or if a CA procedure suffers from a flaw, the records collected by CAs cannot be used as forensic material.
This can basically be solved by mandating DNSSEC when using CAA and also obligate CAs to record DNSSEC chains along CAA records.%

\section{CAA Deployment in the Wild}
\label{sec:method}
We want to assess how the conceptional weaknesses of CAA reflect in operational practice.
To measure CAA, we introduce a novel approach based on CT logs.
Our measurements mimic what we expect from a CAA~Evaluator.
They require strict strict accuracy and flexibility.
Our method aims at \one analyzing as many certificates as possible for each trusted CA, and \two reducing the time discrepancy between certificate issuance and the collection of corresponding CAA records.
For reproducibility reasons, we restrict our measurements to public data, which prevents us from using private DNS zone files.

\paragraph{Certificate Collection}
Recent work confirmed that all major CAs submit their certificates to at least one CT log~\cite{thlsw-dccdi-24}.
Compared to active TLS scans (\eg using top lists), scanning CT logs provide a more complete picture of issued certificates as it also covers certificates which are not actively deployed on any server, \eg backup certificates. %
We scan all active CT logs compliant with the Chrome CT policy.
In June~2024, 10 logs operated by 5 different companies were active.
For each CT log, we regularly check for new precertificates and fetch these.
We collect precertificates instead of final certificates since storing precertificates is mandatory whereas final certificates are optional.

\paragraph{Timely DNS queries}
Maintaining the same set of CAA records as observed by CAs at the time of issuance is challenging because certificates are not instantly available in CT logs after submission.
When a certificate is merged and made publicly available depends on the maximum merge delay (MMD).
Without access to historic DNS records, we argue that this remains the best effort in reconstructing the same CAA records as observed by the issuing~CA.

For all new certificates, we extract all subject alternative names (SAN) of type DNS and expand each domain name to all its parents up to TLDs.
For example, \texttt{www.example.co.uk} is expanded to \texttt{\{www.example.co.uk, example.co.uk, co.uk, uk\}}.
We then query CAA records for all names and store only certificates with at least one SAN that has a corresponding CAA record.
The result is a mapping of domain names to a set of CAA records as well as several other attributes such as the source of record (\eg parent domain), provided the source is a canonical name.

To avoid stale data from third-party caches, 
we retrieve data directly from authoritative name servers instead of recursive resolvers.

\paragraph{Mapping between CA identifiers and CA names}
To verify whether a certificate was issued by a legitimate CA, we need a mapping of CA identifiers (as denoted in CAA records) to CA names (as denoted in X.509 issuer field).
Prior work~\cite{thlsw-dccdi-24,schgn-flcaa-18} and the Mozilla~Common CA~Database
provide such a mapping.
This data, however, is neither up-to-date nor complete.
We extend the Mozilla list by inspecting CPS documents from various~CAs.

Next, for each certificate, we consider all domain names included in the SAN and check whether CAA records comply with the CA that issued the certificate.
Note that in contrast to prior work, our point of departure are certificates and \emph{all} the included domain names and not a single domain name and the corresponding certificate deployed at a web server.

\paragraph{Measurement Setup and Data Corpus}
We queried CT logs over a period of 2 week (2024-06-13 to 2024-06-26) and collected more than $5.4M$ certificates.
About $85\%$ (total of $4.6M$) are unique certificates, because a single cert is logged in multiple logs as required by browsers.
Among multiple instances of the same certificate, we choose the one that was first seen in any of the logs since it represents freshness.
From these certificates we extract $5.8M$ unique domains, of which $34\%$ are wildcards.

\section{Potential Design and Operational Pitfalls}
\label{sec:op-pitfalls}
CAA aims to be cost-effective through low-barrier deployment requirements and simple validation processes~\cite{RFC-8659}.
This is true compared to DNSSEC DANE, MTA-STS, and other DNS-based security protocols.
Such \emph{simplicity}, however, can be delusive as identified in prior work~\cite{schgn-flcaa-18} and discussed among CA operators.
In the following, we discuss pitfalls caused by concessions to accessibility and simplicity of the protocol, investigate how these pitfalls are reflected in current CAA deployment, and propose solutions.

\subsection{Implicit Semantics}
\label{subsec:design-impl-sem}
\paragraph{Problem}
The CAA specification allows to express different semantics using the same syntax or implementing the same semantic using different syntaxes.
For example, \tagI can be used to \one constrain issuance if its value is a CA-recognized domain name, \two forbid issuance if its value is empty (\ie \texttt{";"}), malformed (\eg \texttt{"\%\%\%\%\%"}), or is not associated with any CA, or \three constrain issuance for wildcard names if \tagIW is missing.
Another example is the case of unrestricted issuance by omitting CAA records altogether \emph{if} none of the parents declare CAA records, or by only listing a tag other than \tagI or \tagIW with an arbitrary value.
It is also allowed to have contradicting CAA policies in a CAA set.
For example, to forbid issuance using an empty \tagI tag alongside of another \tagI tag with a valid value.

Such implicit semantics pose two challenges.
First, it hinders implementing specific cases, for example defining constraints only for FQDN names, while posing no restrictions on wildcards.
Second, it prevents CAs from differentiating between intentional configurations and accidental misconfiguration (\eg by typos).
Consequently, definite feedback cannot be provided.%

\paragraph{Deployment}
We observe 811 cases that permit certificate issuance and at the same time contain an empty \tagI tag.
In additional 18 cases, malformed values 
coexist with valid identifiers.
In other cases, we find the same erroneous, yet syntactically correct, values used at different domain names, indicating that the name owners have copied the values without realizing its semantics.
For instance, we observe 1,319 unique domain names using \texttt{0issueletsencrypt.org} (incorrectly includes flag and tag within the value) with 829 using the following exact set to issue constraints: \texttt{\{0issueletsencrypt.org, comodoca.com, digicert.com, letsencrypt.org, pki.goog\}} (note the correct value for Let's Encrypt).
Although the CAA specification explicitly ignores such cases~\cite{RFC-8659}, specifically for values that hint at a typo rather than an intentional malformed value, we diagnose that the semantic is not fully understood by name owners and can lead to unwanted results.

In our dataset, there are also cases of unintentional lifting of constraints due to unknown tags.
Two domains use \texttt{isssue} tag (note extra ``s'') with a value of \texttt{letsencrypt.org}, which evidently was meant to constrain issuance, but in practice lifts all restrictions.
Another domain uses the aforementioned erroneous tag alongside the correct \tagI tag.
Similarly, there are cases where \tagI was misspelled (\eg \texttt{issed}) next to a valid \tagIW records, thus  only limiting issuance for wildcard domains.
We also observe confusion by an issuer parsing the CA identifier only in \tagIW but not the \texttt{issue} tag, leading a CA to consider itself to be authorized to certify FQDNs despite CAA constraints.
We reported this incident to the CA (GoDaddy), which confirmed the failure (see \href{https://bugzilla.mozilla.org/show_bug.cgi?id=1904748}{Bugzilla \#1904748}).

\paragraph{Solution}
First, the semantics of each tag must be uniquely confined, \ie \tagI is limited to FQDN domains, \tagIW only applies to wildcard domains, and existence of (or lack thereof)  one does not impact the other.
Second, a constant value should be standardized to denote a no-issue policy instead of empty or malformed values.
This way typos (as discussed below) or other unintentional errors are explicitly detected and  not open to  misinterpretation.
Finally, contradictory policies should be considered as errors.

\subsection{Boundless Policy Scoping}
\label{subsec:problem-boundless}
\begin{figure}
    \includegraphics{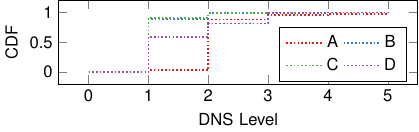}
    \caption{Distribution of domain names by DNS level at which relevant CAA records were found. Group A are delegated domains w/o CAA, B are alias domains w/o CAA, C are alias parents w/ CAA, and D are alias domains w/o CAA but with an alias parent with CAA.}
    \label{fig:grouped-trans-auth-caa-depth}
\end{figure}

\paragraph{Problem}
The DNS namespace is partitioned in zones.
Each zone comprises a set of authoritative data for all the names in its namespaces.
A zone owner can further delegate parts of his namespace to others and create new zones.
Delegated zones are independent of their parents so that the parent zone cannot define records for names in a delegated namespace without retracting the delegation.

The CAA specification implicitly breaks this authoritative barrier by allowing CAA records to be inherited from \emph{parents}
 in a different authority.
Consequently, the attack surface is expanded beyond the owned name servers.
For instance, if parent and child define contradictory policies, or there is a parent applying less strict policies, an attacker can try to suppress or spoof all CAA queries in the query chain until the desired level is reached. Also the attacker could compromise a name server in the hierarchy to impact the entire namespace.
It should be noted that not only constraining policies 
are inherited, but also contact information that are used for domain validation~\cite{bf-brimp-24}, \eg as specified by \caaTag{contactemail} tag.
This allows an entity to be granted a certificate for a domain name which it does not control.

A similar situation arises for aliases,
which do not allow other records to be present except DNSSEC crypto. 
Here, a CA first asks the authority of the canonical name for CAA records, implying that only a potentially different authority can define CAA policies for a domain name that is not delegated.
The zone owner is left without any means to control CAA certification control.

\paragraph{Deployment}
More than $777k$ domain names in our dataset are subject to CAA policies defined by a separate zone.
We classify these into four categories (ordered by frequency of occurrence):

\begin{description}
	\item[A] \textit{Delegated domain w/o CAA.} In more than half of all cases, the CAA is defined by a parenting zone for a subdomain that is delegated.
	\item[B] \textit{Alias domain w/o CAA.} About $ 45\%$ of the cases show alias domains with CAA records sourced from a parent, \ie the authority of the domain name and CAA source differ.
	\item[C] \textit{Alias parent w/ CAA.} 1,672 instances inherit CAA and the relevant CAA records originate from a parenting domain name which is an alias for a domain name in another zone.
    \item[D] \textit{Alias domain w/o CAA and alias parent w/ CAA.} As a special instance of the former case,  the domain name is itself an alias without CAA records, but we find a parent that is also an alias and its canonical name (from another zone) provides effective CAA records ($1,080$ instances).
\end{description}

\noindent
In total, there were 108 domains for which a parent in another zone lists email addresses.
If the CA supports domain validation using CAA \texttt{contactemail}, the parent can apply and successfully retrieve a domain validated (DV) certificate for such a subdomain in another zone. 
In all instances it is another authority that defines CAA policies for a namespace.
\autoref{fig:grouped-trans-auth-caa-depth} depicts the distribution of DNS hierarchy levels at which  CAA records were found.

\paragraph{Solution}
The effective scope of CAA records must be  confined within a DNS zone.
A na\"ive approach would adapt the algorithm used to find relevant CAA records~\cite{RFC-8659} to stop at the zone apex and not at the DNS root.
This, however, does not solve the problem of aliases.
The conceptually clean solution taken by DANE~\cite{RFC-6698} dedicates subdomains for \texttt{TLSA} records.
Thus in case of canonical names, the referred  authority has no control over DANE records.
Analogously, CAA records should be listed under a designated \texttt{\_caa} subdomain, \eg \texttt{\_caa.example.com}.

\subsection{Ambiguous Identifiers}
\label{subsec:design-amb-id}
\paragraph{Problem}
The value of \tagI and \tagIW tags is called `issuer-domain-name' and identifies the CA which owns that name.
This identification, however, is not unique as the same domain name is also accepted by any `party acting under the explicit authority of the holder of the issuer-domain-name'~\cite{RFC-8659}.
This can be a subsidiary, a reseller, or even another CA organization.
Ambiguous identifiers defeat the purpose of CAA by preventing name owners to authorize a specific CA.
Two cases are plausible.
First,  a name owner unknowingly authorizes a third-party CA.
This happens when independent CAs accept the same identifiers, \eg DigiCert also accepts Amazon identifiers
as it vicariously runs some intermediate CAs for Amazon.
Second, a name owner unknowingly authorizes all CAs running on the same infrastructure.
This is the case for resellers, managed PKIs and alike.
For example, DKB, a German bank, has its own dedicated CA 
managed by DigiCert, but  lacks a dedicated identifier. Hence it uses `digicert.com', which authorizes all DigiCert subsidiaries, such as Thawte or QuoVadis.

\begin{table*}
    \centering
    \caption{The CA strings that appear in more than 10\% of the CAA records in our dataset as well as their occurrence as the \textit{only} CA string.}
    \begin{tabularx}{\textwidth}{X r r r r r}
      \toprule
      & \multicolumn{2}{c}{\tt issue} & \multicolumn{2}{c}{\tt issuewild} & \multirow{3.1}{*}{\shortstack{CA\\Count$^\dagger$}}\\
      \cmidrule(lr){2-3}
      \cmidrule(lr){4-5}
      CAA String               & Overall  & Single     & Overall  & Single & \\
      \midrule
      \texttt{letsencrypt.org} & 89.96\%  & 37.62\%    & 87.26\%  & 5.68\% & 1\\
      \texttt{digicert.com}    & 38.85\%  &  0.40\%    & 87.26\%  & 0.93\% & 3\\
      \texttt{pki.goog}        & 36.21\%  &  5.90\%    & 72.22\%  & 8.46\% & 1\\
      \texttt{comodoca.com}    & 27.84\%  &  0.11\%    & 67.16\%  & 0.03\% & 6\\
      \texttt{globalsign.com}  & 26.47\%  &  0.02\%    & 44.66\%  & 0.12\% & 2\\
      \texttt{sectigo.com}     & 25.00\%  &  0.05\%    & 33.48\%  & 0.21\% & 8\\
      \texttt{amazon.com}      & 11.24\%  &  $<$0.01\% &  1.92\%  & 0.04\% & 1\\
      \texttt{godaddy.com}     &  8.25\%  &  $<$0.01\% &  $<$0.01\%  & $<$0.01\% & 2\\
      \bottomrule
      \multicolumn{6}{l}{\shortstack[l]{$^\dagger$ Count of unique CAs (by Subject Organization) in our dataset that match the respective CAA string.}}
    \end{tabularx}
    \label{tab:ca_string_shares}
  \end{table*}
\paragraph{Deployment}
Without additional information, it is not possible to detect whether name owners are aware of identifier ambiguities.
In our dataset, nearly every fifth CAA source redundantly lists \texttt{comodoca.com} and \texttt{sectigo.com} as values for \texttt{issue}, whereas Sectigo CPS explicitly states that the former identifier is deprecated.
We also observe up to eight different CA organizations that share the same identifier.
Note that the number of CAs sharing identifiers is higher than what we observed in our measurements.
Alone \texttt{sectigo.com} is accepted by Gandi (FR), ZeroSSL (CH), GEANT Vereniging (NL), eMudhra Technologies Limited (IN), and at least 17 other CAs from around the world that use Sectigo infrastructure.

Our data also revealed cases of misissuances (confirmed by the CA GoDaddy), in which all strings matching a pattern were accepted instead of a fixed domain name (see \href{https://bugzilla.mozilla.org/show_bug.cgi?id=1904749}{Bugzilla \#1904749}).
\autoref{tab:ca_string_shares} gives an overview of the most frequently used identifiers and shows how many organizations match the same identifier.

\paragraph{Solution}
The identifiers should reference the CA with whom subscribers directly enter a business relationship instead of the infrastructure operator.
Furthermore, an authoritative and publicly accessible mapping of CAA identifiers to CAs must be provided so that \one name owners can verify the reach of each identifier, and \two any other identifier be considered  invalid.
Although Mozilla maintains such a mapping in its `Common CA Database',
it is neither authoritative nor complete~\cite{schgn-flcaa-18}.
The protocol registry of Internet Assigned Numbers Authority (IANA) is a potential alternative to maintain such a mapping, as was proposed for CAA tags~\cite{RFC-8659}.

\subsection{Non-verifiable and Temporally Unbound Policies}
\paragraph{Problem}
The CAA specification describes Evaluators as third-party auditors that may use CAA records to detect policy violations~\cite{RFC-8659}.
At the same time, it warns that DNS records may have changed at the time an Evaluator retrieves them and are effectively unreliable in discovering misissuances.
In its current state, CAA can only be verified by the name owner and the CA at the time of certification.
Even at that time, there is no definitive way for the CA to make sure that CAA records have not been manipulated or suppressed on-the-fly.
This issue has its roots in missing \one validity timestamps, and \two data-origin and integrity verification features.

\begin{figure}
    \includegraphics{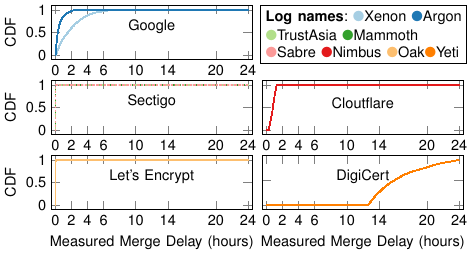}
    \caption{Distribution of measured CT merge delays}
    \label{fig:ts-diff-ecdf}
\end{figure}

\begin{figure}
    \includegraphics{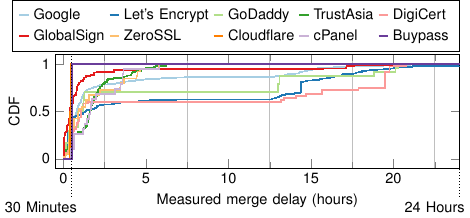}
    \caption{Distribution of measured merge delay for certificates with at least one mismatching CAA record}
    \label{fig:ts-diff-mismatch-ecdf}
\end{figure}
\paragraph{Deployment}
To the best our knowledge, there are no publicly active CAA Evaluators.
Using the measurements in this paper, we take the role of an Evaluator.
Our proposed method in \autoref{sec:method} reduces the temporal gap between CA query of CAA records and our own queries.
Nonetheless, a time discrepancy as large as the maximum merge delay of CT logs (see \autoref{sec:method}) remains.
\autoref{fig:ts-diff-ecdf} depicts this merge delay for various CT logs as measured by our scanners.

To understand the impact of time discrepancy we take at look at cases of mismatching the CAA policies in our dataset.
Given that \one we observe only a small fraction of certificates that mismatch their corresponding CAA records, and \two we are unable to reproduce such cases independently, we conjecture that these are false negatives due to modified DNS records.
\autoref{fig:ts-diff-mismatch-ecdf} shows the distributions of these cases with respect to the measured merge delay.
We can conclude that any time discrepancy larger than 30 minutes significantly increases the chance of incorrect CAA validation.
Regardless of merge delays, CAs can also use CAA records as stale as 8 hours~\cite{bf-brimp-24}, so that in the worst case the temporal discrepancy amounts to $8h + MMD$.
Even \href{https://sunlight.dev/}{Sunlight logs}, a new generation of CT logs that remove merge delays, cannot fill this gap.
Without fine-granular historical DNS data or designated validity timestamps for CAA records, auditing of CAA remains unreliable.

\paragraph{Solution}
This problem can be addressed indirectly by enabling DNSSEC or directly by adding validity timestamps to CAA records and signing them.
The former approach infers validity of a record from validity period of covering DNSSEC signatures,
while data-origins can be validated using DNSSEC keys for that zone.
In the latter approach CAA records are extended to \one carry a validity timestamp, and \two be cryptographically secured, \eg through a digital signature.
This also allows to determine which entity defined a set of CAA policies (data-origin authentication).
A signed and temporally constrained CAA record can retrospectively and securely be verified, \eg by an Evaluator.

\section{Lessons learned}
\label{sec:disc}
CA Authorization was designed as a preventive measure against certificate misissuance but falls short due to design decisions, which hinder proper deployment.
We highlighted pitfalls and proposed solutions.
In the following, we report about lessons learned from a decade of CAA usage and discuss aspects that should guide the design of related protocols in the future.

\paragraph{DNS is a fragile basis for security protocols}
Any service that relies on the unprotected DNS inherits its vulnerabilities.
For example, it has been shown how plain DNS can impact domain validation and consequently lead to certificate misissuance~\cite{sww-difsd-19}.
Although aware of this issue, CAA does not provide an alternative and only recommends using DNSSEC
---without making it mandatory---to mitigate threats targeting DNS~\cite{RFC-8659}.
DNSSEC, a set of security extensions, brings integrity and origin authentication alongside authentication of name nonexistence to DNS.
Consequently, DANE mandates the use of DNSSEC.
DNSSEC, however, is not widely deployed by name servers and resolvers~\cite{ormz-qosdd-08,cvcclmmw-trbdbac-2017,otsw-fbtfy-22}.

Solutions orthogonal to DNSSEC have been developed to mitigate DNS security weaknesses, among them Multi-Perspective Issuance Corroboration (MPIC), and trust on first use (TOFU). MPIC verifies domain control from multiple vantage points across the Internet to reduce the risk of retrieving manipulated data.
It has  been
integrated
into CA/Browser Forum Baseline Requirements as part of domain validation procedures. 
TOFU-based approaches rely on a first intact and uncompromised data exchange to establish security parameters for upcomming interactions.
An example is MTA-STS,
which is used against downgrade attacks in STARTTLS protocols (\eg deployed by mail servers). 

\paragraph{Trust must be derived from frequent public audits}
To measure effectiveness of a security protocol, it must be auditable.
Experiences from the Web PKI show that formal and private audits (\eg WebTrust or ETSI) are rather blunt tools for preventing or detecting certificate missiuances.
In contrast, providing the public with adequate means to perform  audits is  effective as the success story of CT logs reveals.

CAA validation by CAs cannot be audited.
CAs do not publicly log CAA record consumption nor even provide them upon request~\cite{schgn-flcaa-18}, 
and even if they did, they could not rule out manipulation or spoofing.
For DNS-based protocols to be auditable, some form of content object security is required for the underlying resource records to be temporally and spatially decoupled from the name owner---as provided by DNSSEC.
Once object security is reached for DNS resource records, decoupled data can be distributed over established  channels, for example as extensions to the Online Certificate Status Protocol (OCSP)
or  TLS (\cf TLS DNSSEC Chain Extension).

\paragraph{Potential victims must be able to verify measures}
CAA has been designed as a light-weight approach to protect against certificate misissuance. It enables  the CAs---and only the CAs---to verify their procedures at the time of certificate issuance.  

Potential victims of illegitimate certificates, however, are resource owners and service users. An online financial institution, for example, looses reputation if its customers get fooled by an illegitimately certified fraud service. Neither the legitimate service provider nor its users can reliably detect cases of misissuance, which leaves them blind with respect to the efficacy of the CAA protective measures. This blindness of potential victims turns CAA into a toothless security scheme. Instead, CAA merely serves as a detection aid for configuration errors in  toolchains of well-behaving CAs.

\vspace*{-10pt}
\section{Conclusions and Outlook}
\label{sec:conclusion}

A reliable and trustworthy certificate issuing process is utmost important to enable secure Internet services.
In this paper, we critically revisited CAA, a currently popular DNS-based Internet standard for preventing incorrect certificates. The CA/Browser Forum mandates the deployment of CAA.
We found the design of CAA flawed to an extent that  prevents proper deployment and proper auditing.
Given the importance of the security objectives and that some problems have been found in the wild already in 2018, the lack of progress leaves us puzzled.

We identified several options that require only minor changes to improve CAA in future development, and highlighted principle design choices that should guide the design of improvements.
With these insights we hope to have shed light on how to build a more trustworthy certificate~ecosystem in the near future.

\paragraph{Ethical Concerns}
This position paper draws attention to a security approach that should be improved with respect to design and deployment.
When we found incidents in real deployments, we implemented a responsible disclosure policy to solve the problem with the concerned CA before making the incidents public.

\vspace*{-15pt}
\section{Acknowledgments}
This work was supported in parts by the German Federal Ministry of Education and Research (BMBF) within the project PRIMEnet.

\label{lastpage}

\vspace*{-5pt}
\bibliographystyle{IEEEtran}
\bibliography{attacks,internet,ids,rfcs,security,own}

\newpage

\begin{IEEEbiography}{Pouyan Fotouhi Tehrani}{\,} is a final-year PhD student and a research associate in the Chair of Distributed and Networked Systems at TU Dresden. His research focuses on communication networks for emergencies and crises, particularly object security in fragmented and challenged networks.
\end{IEEEbiography}

\begin{IEEEbiography}{Raphael Hiesgen}{\,} is currently pursuing a Ph.D. at HAW Hamburg under the guidance of Prof. T. C. Schmidt and Prof. Wählisch. His research centers on Internet measurements, specifically in the context of security applications. Proficient in C++ and adept at crafting scalable systems, he developed Spoki, a reactive network telescope tailored for long-term measurements. His skill set spans from designing and implementing measurement systems to delving into the analysis of collected data. Spoki was initially presented at USENIX Security in 2022 and is undergoing continuous improvements to broaden its visibility into the scanning ecosystem.
\end{IEEEbiography}

\begin{IEEEbiography}{Thomas C. Schmidt}{\,}
Thomas C. Schmidt is a Professor of computer networks and Internet technologies with the Hamburg University of Applied Sciences, Hamburg, Germany, where he heads the Internet Technologies Research Group.
He was the Principal Investigator in a number of EU, nationally funded, and industrial projects, as well as a Visiting Professor with the University of Reading, Reading, U.K. 
He is also a co-founder and a coordinator of the open source community developing the RIOT operating system. 
His current research interests include development, measurement, and analysis of large-scale distributed systems like the Internet or its offsprings. 
\end{IEEEbiography}

\begin{IEEEbiography}{Matthias W\"ahlisch}{\,}
Matthias W\"ahlisch holds the Chair of Distributed and Networked Systems at TU Dresden, and is a Research Fellow of the Barkhausen Institut.
His research and teaching focus on scalable, reliable, and secure Internet communication.
This includes the design and evaluation of networking protocols and architectures, as well as Internet measurements and analysis.
Matthias is involved in the IETF since 2005 and co-founded multiple successful open source projects.
\end{IEEEbiography}

\end{document}